\documentclass[journal=jpccck,manuscript=article,layout=traditional]{achemso}

\usepackage{graphicx}
\usepackage{dcolumn}
\usepackage{bm}
\usepackage{times}
\usepackage{xcolor}
\usepackage{natmove}
\usepackage{multirow}
\usepackage{setspace}
\usepackage{caption}
\usepackage{array}
\usepackage{booktabs}
\usepackage{rotating}
\usepackage{longtable}
\usepackage{amsmath}
\usepackage{achemso}
\setkeys{acs}{usetitle = true}
\setkeys{acs}{maxauthors=10, etalmode=truncate}
\usepackage[capitalise,noabbrev]{cleveref}
\usepackage{color}

\title{FePc Adsorption on the Moir\'e Superstructure of Graphene Intercalated with a Co Layer}

\author{Giulia Avvisati}
\affiliation{Dipartimento di Fisica, Universit$\grave{a}$ di Roma ``La Sapienza'', I-00185 Roma, Italy}
\author{Simone Lisi}
\affiliation{Dipartimento di Fisica, Universit$\grave{a}$ di Roma ``La Sapienza'', I-00185 Roma, Italy}
\author{Pierluigi Gargiani}
\affiliation{ALBA Synchrotron Light Source, E-08290 Cerdanyola del Valles, Barcelona, Spain}
\author{Ada Della Pia}
\affiliation{Dipartimento di Fisica, Universit$\grave{a}$ di Roma ``La Sapienza'', I-00185 Roma, Italy}
\author{Oreste De Luca}
\affiliation{Dipartimento di Fisica, Universit$\grave{a}$ della Calabria, I-87036 Arcavacata di Rende (CS), Italy}
\author{Daniela Pacil\'e}
\affiliation{Dipartimento di Fisica, Universit$\grave{a}$ della Calabria, I-87036 Arcavacata di Rende (CS), Italy}
\author{Claudia Cardoso}
\affiliation{Centro S3, CNR-Istituto Nanoscienze, I-41125 Modena, Italy}
\author{Daniele Varsano}
\affiliation{Centro S3, CNR-Istituto Nanoscienze, I-41125 Modena, Italy}
\author{Deborah Prezzi}
\affiliation{Centro S3, CNR-Istituto Nanoscienze, I-41125 Modena, Italy}
\author{Andrea Ferretti}
\affiliation{Centro S3, CNR-Istituto Nanoscienze, I-41125 Modena, Italy}
\author{Maria Grazia Betti}
\affiliation{Dipartimento di Fisica, Universit$\grave{a}$ di Roma ``La Sapienza'', I-00185 Roma, Italy}
\email{maria.grazia.betti@roma1.infn.it}

\begin{document}

\begin{abstract}
\noindent
 The moir\'e superstructure of graphene grown on metals can drive the assembly of molecular architectures, as iron-phthalocyanine (FePc) molecules, allowing for the production of artificial molecular configurations. A detailed analysis of the Gr/Co interaction upon intercalation (including a modelling of the resulting moir\'e pattern) is performed here by density functional theory, which provides an accurate description of the template as a function of the corrugation parameters. The theoretical results are a preliminary step to describe the interaction process of the FePc molecules adsorption on the Gr/Co system. Core level photoemission and absorption spectroscopies have been employed to control the preferential adsorption regions of the FePc on the graphene moir\'e superstructure and the interaction of the central Fe ion with the underlying Co. Our results show that upon molecular adsorption the distance of C atoms from the Co template mainly drives the strength of the molecules-substrate interaction, thereby allowing for locally different electronic properties within the corrugated interface.
\end{abstract}

\maketitle

\newpage
\section{Introduction}

Equally-sized and evenly-spaced molecular networks can be active centers for electronic and magnetic functionalities. Organo-metallic molecules can be excellent building blocks of ordered 1D and 2D structures assembled on suitable templates, as they are involved in weak directional bonds, and can easily form supramolecular highly-ordered architectures~\cite{Lin2009, Fortuna_JPCC_2012}. The choice of the template surface is strategic to drive 1D~\cite{Betti_Langmuir_2012} and 2D~\cite{HamalainenJPCC2012, Barrena_ChemPhysChem_2007, Scardamaglia2011} long range ordering, and to control the interaction among the molecular units. Corrugated epitaxial graphene (Gr) on metal surfaces has been used to host ordered molecular superstructures~\cite{MacLeod_SMALL_2014, Barja_ChemComm_2010}, thanks to moir\'e lattice super-periodicity (of the order of 1.5-3 nm)~\cite{Wintterlin20091841}. These 2D-supported networks represent ideal systems since the molecular lattice can be stable in a wide temperature range, with long-range ordering extended over the whole Gr layer (several microns) and tunable coupling, depending on the commensurability, with the metal substrate~\cite{PhysRevB.78.073401}. The fine control of the corrugation and/or interaction strength is a strategic route to design novel ordered molecular superstructures with desired functionalities. 

 Recently, the intercalation of transition metals between Gr and Ir(111) has been proposed~\cite{Pacile2014, VitaPRB2014, BazarnikACSnano2013} as a valuable route for tuning the self-assembly of organometallic molecules. The surface potential of the Gr layer is modified via intercalation, leading to a moir\'e superlattice with higher height modulations.  Molecules can then adsorb in preferential sites with the formation of different 2D superstructures, driven by the  Gr corrugation and the interaction strength with the intercalated metal. The geometric and electronic modulation at the nanometer scale of the moir\'e can be amplified by cobalt (iron, nickel) intercalation.  Highly corrugated graphene substrate (1.2-1.8 {\AA} for Co~\cite{PhysRevB.87.041403}, 1.3 {\AA} for Fe~\cite{Decker_JPhys_2014}) induces a strong modulation of the surface potential and it can drive the assembly of highly ordered arrays, e.g. Kagome superlattices, as observed for Metal-Pc assembled on Gr/Ru~\cite{JAmChemSoc.131.14136} and Gr/Co(Fe)/Ir~\cite{BazarnikACSnano2013}. Structural control of the rippling depends also on the number of intercalated layers. In particular, the first Co epitaxial layer is arranged in registry with the Ir(111) surface lattice and the stronger electronic interaction induces an enhancement in the moir\'e corrugation, going from 0.35 {\AA} up to 1.2-1.8 {\AA}, as shown by a recent Scanning Tunneling Microscopy (STM) study~\cite{PhysRevB.87.041403}. Further intercalation of Co layers releases the Gr-Co mismatch,~\cite{Pacile2014} as the topmost Co layers can partly recover the bulk value of the planar lattice constant (2.42 {\AA}), which is almost commensurate with that of freestanding Gr (2.44 {\AA}).~\cite{khom+09prb}
 Moreover, the first Co intercalated layer shows a magnetic anisotropy with an out of-plane easy magnetization axis~\cite{PhysRevB.87.041403, VitaPRB2014}, in contrast to its bulk configuration. The combination of these magnetic properties with strong corrugation makes this substrate an ideal template for growing organo-metallic nano-architectures, with the final aim at designing low dimensional nano-scaled spintronic devices.
 
 In this paper, we first perform a theoretical investigation, based on density functional theory (DFT), of the interaction between Gr grown on Ir(111) and an intercalated layer of Co, with special attention to the comparison with experimental photoemission data (X-ray Photoemission Spectroscopy, XPS). This first step is preparatory to understand the assembly and the interaction process of FePc at increasing molecular density. A detailed understanding of the role of the macrocycle ligands and of the Fe metal ions in the interaction process, as a function of the adsorption sites, is addressed by means of core level photoemission and absorption spectroscopies. Our findings show that the FePc adsorption on the valley regions of the moir\'e superstructure is favoured in the submonolayer regime, while the hills are filled lastly. Accordingly, a strong Gr-mediated electronic interaction between Fe atoms and the Co interface is revealed at low molecular coverage.

\section{Results and Discussion}

In this section first we discuss how the interaction strength (as measured by C chemical shifts) of Gr/Ir intercalated with Cobalt is modified by the corrugation due to the moir\'e patterning. 
Next, based on these results, we present and discuss the interaction of FePc on Gr/Co, as measured by XPS and NEXAFS.

\subsection{Graphene interaction with intercalated Co: a theoretical perspective}
\label{sec:results_XPS_theo}

\begin{figure}
\centering
\includegraphics[clip,width=0.50\textwidth]{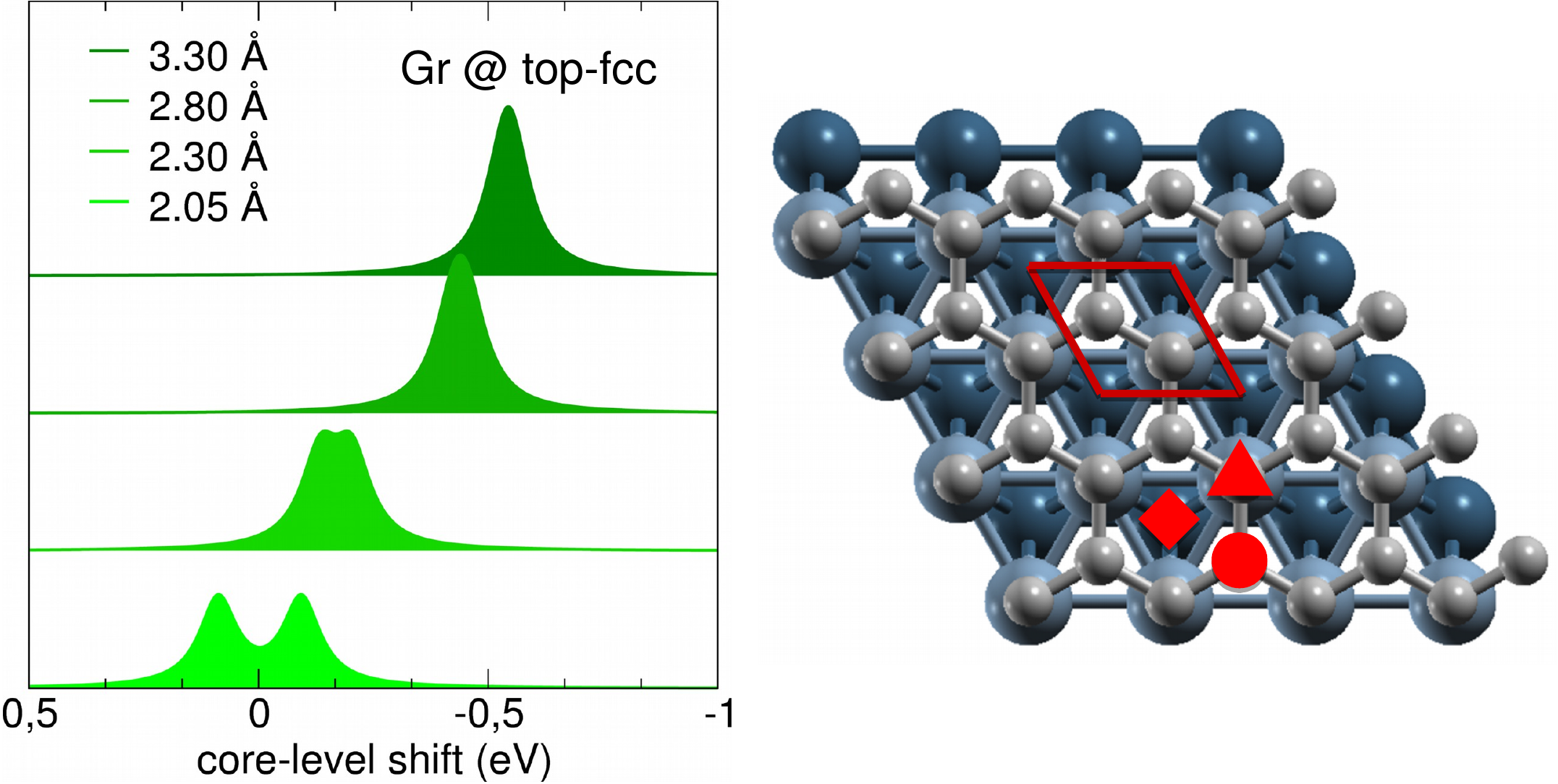}
\caption{Left panel: C$1s$ core-level shifts computed for the two non-equivalent C atoms of Gr@Co(0001) in top-fcc registry (C atoms at fcc-hollow and on-top sites; through the paper, registries are named according to the position of the two Gr atoms on the hpc surface), with increasing graphene-Co distance and represented as Lorentzian functions with a width of 0.1~eV. The average of the CLS computed for GR@top-fcc at 2.05\AA~distance was used as reference, here set to zero. Right panel: top view of the studied fcc-top Gr@Co(0001) geometry with the grey representing C and blue Co atoms. The top Co layer is represented in a lighter blue. The 1$\times$1 unit cell is reported. Circle and triangle symbols refer to {\it fcc} and {\it top} adsorption sites, respectively. Diamond symbols correspond to the {\it hcp} site.
\label{fig:cls_fcc}}
\end{figure}

\begin{figure}
\centering
\hspace{0.08\textwidth}\includegraphics[clip,width=0.29\textwidth]{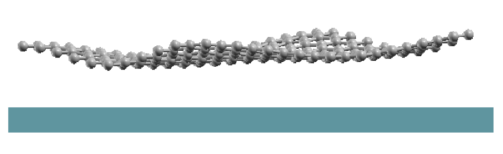}\\
\includegraphics[clip,width=0.49\textwidth]{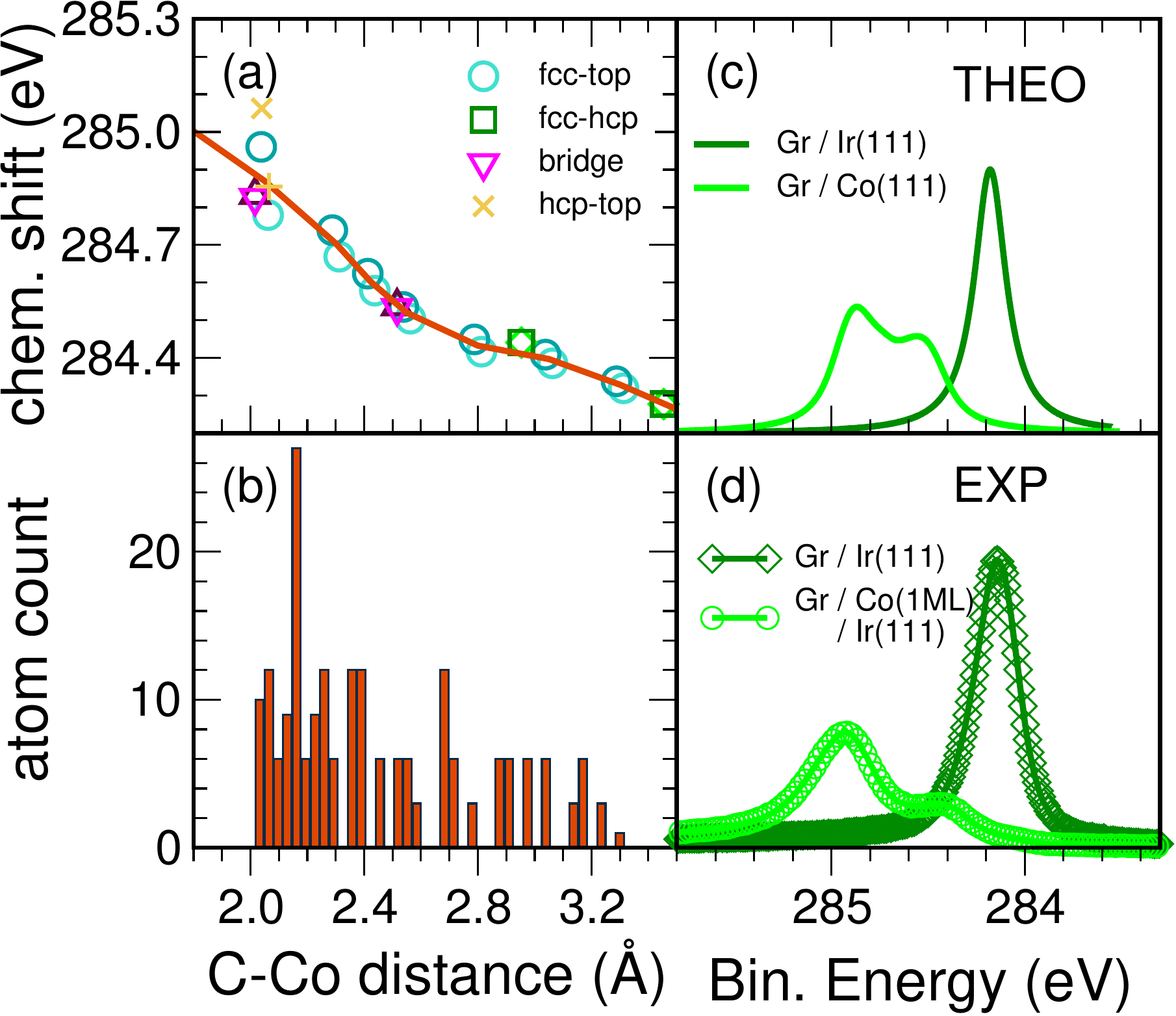}
\caption{
Top panel: a cartoon of the corrugated Graphene as given by the egg-box model.
(a) Core-hole shift dependence with C-Co distance computed for Gr@Co. The red line gives the average of the values computed for the two non-equivalent C sites of four different registries of commensurate Gr@Co, shown by the symbols in different colours.
(b) Height distribution for corrugated Gr@Co as given by the egg-box model~\cite{PhysRevLett.107.036101}. 
(c) XPS spectra computed as a sum of Lorentzian functions centered in the core-hole energies given in the left panel.
(d): Experimental C$1s$ XPS spectra ~\cite{Pacile2014}.\label{fig:cls_DFT}}
\end{figure}

The chemical shifts of carbon atoms for Gr/Ir intercalated with a Co monolayer have been recently studied by means of photoemission spectroscopy~\cite{Pacile2014,VitaPRB2014}. The XPS spectra (also reported in Fig.~\ref{fig:cls_DFT}d) show a double peaked line profile, with components at 284.92 and 284.42~eV. When one Co layer is intercalated between Gr and Ir, graphene shows a characteristic moir\'e pattern due to the lattice mismatch with a pronounced corrugation (the peak-valley modulation of C atoms is experimentally estimated to be 1.2-1.8{~\AA}~\cite{PhysRevB.87.041403}).
Both corrugation and hybridization of the graphene layer are expected to influence the C$1s$ core level line shape, as observed for other corrugated graphene layers grown on selected metallic surfaces, like Ru(0001), Rh(111), and Re(0001)~\cite{PhysRevB.78.073401,Miniussi2011,Alfe2013,Papagno_PRB_2013}. The double component of the XPS spectra can be associated to a corrugated graphene layer, where the peak with lower binding energy is due to carbon atoms weakly interacting with the surface, while the high-energy peak results from highly interacting carbon atoms.
Detailed DFT calculations on Gr/Ni/Ir(111)~\cite{Pacile2013} and Gr/Re(0001)~\cite{Alfe2013}
interfaces have been employed to simulate structural properties and XPS spectra and they further support this interpretation of the two XPS peaks observed experimentally.
In order to better understand the origin of the two-peaked structure of XPS spectra obtained from Co-intercalated Gr/Ir, we have also performed ab initio calculations based on DFT.

The description of graphene corrugation requires a large unit cell of 200 C atoms plus the underlying Co surface. After the geometry optimization of such a large system, the XPS spectra should then be determined using the core-level shifts computed for each of these C atoms.
Based on previous results for Gr/Re(0001)~\cite{Alfe2013}, we began by studying the dependence of the C$1s$ chemical shifts on the C-Co distance and relative registry, neglecting the moir\'e corrugation of graphene, which permits the use of smaller unit cells. Then we combine this information with a model for the moir\'e patterning which permitted to simulate the actual XPS C$1s$ energy splitting without explicitly performing a calculation for each of the 200 C atoms. The smaller unit cells and the smaller number of calculations required, lowered considerably the computational cost for the XPS spectra calculation.

\subsubsection{XPS of flat graphene @ Co}

We start with the study of graphene adsorbed flat on Co(0001). As discussed in the literature~\cite{khom+09prb}, there is only a small mismatch (smaller than 1\%) between Co bulk and free standing Gr lattice constants (2.42 and 2.44~\AA, respectively, according to our calculations, and in good agreement with existing data~\cite{khom+09prb}). Graphene is then found to grow flat on Co(0001) with a commensurate lattice~\cite{PhysRevX.2.041017,khom+09prb,Pacile2014}. Among several possible registries, the one in which graphene occupies the so-called top-fcc positions has been shown to be the most stable~\cite{prez+14acsnano}. In this registry, C atoms occupy two non-equivalent sites of the hcp Co(0001) surface, one {\it on-top} and the other {\it fcc-hollow}, as illustrated in the right panel of Fig.~\ref{fig:cls_fcc}. From the experimental point of view, the existence of two main components in the C$1s$ peak of Gr adsorbed flat on Co (energy splitting of about 100 meV) was recently proved by monitoring the linewidth modulation, as a function of the emission angle~\cite{Pacile2014}. More recently, by photoelectron diffraction measurements performed on the same system, a separation value between the two main components of about 270 meV has been estimated~\cite{Usachov2016}.

At the DFT-LDA level, the optimized Co-Gr distance for this configuration is found to be 2.05~\AA{}, in good agreement with existing literature~\cite{khom+09prb,prez+14acsnano}.
The C$1s$ core-level shifts computed for these two C sites differ by about 180~meV. The splitting between these two peaks had been previously computed within the {\it initial states approximation} in Ref.~\cite{Usachov2016} (ISA), where a larger value of 350~meV was found. This discrepancy can be explained by the fact that ISA neglects relaxation effects of the valence and core electrons upon ionization of the core. In particular, within ISA, the binding energy of the photo-emitted core electron is approximated, in a Koopmans-like approach, by minus the core level eigenvalue.
Instead, the method that we have used, the core-hole $\Delta$SCF approach, takes the opposite limit and assumes that the valence electrons fully relax to the presence of the hole created at the core during the photoemission process. The core level energies are then computed as a difference of total energies between the initial and final states of the system, in which the valence electrons are in their ground state both before and after the core electron removal. \\
A full C-Co distance dependence of the C$1s$ core level shifts is shown in Fig.~\ref{fig:cls_fcc}(a), which proved to be largely independent of the registry (see SI).
Interestingly, the shift towards lower binding energies with increasing distance has been found to be definitely larger than the one due to non-equivalent sites. For instance, the energy shift for the average of the two core level shifts computed at 2.05~\AA{} and at 3.30~\AA{} Gr/Co distances is 0.54~eV. \\

\subsubsection{Modelling the moir\'e pattern}

In order to model the corrugation of graphene on top of the Co surface we considered a 10$\times$10 graphene supercell and used the so called egg-box model in which the heights of the C atoms are given by~\cite{PhysRevLett.107.036101}
\begin{equation}
h(\mathbf{r})= h_{\text{avg}} + \frac{2}{9} \Delta h \,
\left[ \sum_{i=1}^3 \cos(\mathbf{k}_i \cdot \mathbf{r})
\right]
\label{eq:moire}
\end{equation}
where $\Delta h = h_{\text{max}}-h_{\text{min}}$, $h_{\text{avg}} = 1/2(h_{\text{max}}+h_{\text{min}})$, $|\mathbf{k}_i|=k$, and $\angle \mathbf{k}_i \mathbf{k}_j =120^\circ$ for $i\neq j$.
For the maximum and minimum distances between C and the Co surface we used the values published in Ref.~\cite{PhysRevB.87.041403} ($h_{\text{max}}$=3.29~\AA, $h_{\text{min}}$=2.02~\AA), where a complete geometry optimization of the 10$\times$10/9$\times$9 Gr/Co/Ir supercell was performed using DFT at the GGA+U level and including van der Waals corrections. Even if we use a different DFT functional (LDA),
these values are in good agreement with the distances that we obtained for a commensurate cell for different Gr/Co registries. Indeed, according to the moir\'e pattern of Gr@Co measured by STM, the regions corresponding to the largest C-Co distances show a Gr/Co registry in which the center of the graphene hexagons are on top of the Co atoms (fcc-hpc registry in our notation). The computed distance for the commensurate Gr@Co cell with the same registry is 3.45~\AA{}. On the other hand the fcc-top and hpc-top registries are shown to be present in the valleys of the corrugated graphene sheet, and match rather well with the distance computed for commensurate Gr@Co cells (2.05 and 2.04~\AA{}, respectively). This suggests that the main driving force for the corrugation of the moir\'e pattern is the actual local Gr/Co registry~\cite{prez+14acsnano}.

The main outcome of the above model is the distribution of the C-Co distances in the presence of the moir\'e corrugation, which is plotted in Fig.~\ref{fig:cls_DFT}(b). We then used the core-level shifts computed for the commensurate Gr@Co at different distances and for the different registries to determine the core-level shifts for each of the C atoms in our model. For larger Co-Gr distances, the core level shifts are mostly independent on the registry and on the C sites. In the case of the shortest Co-Gr distances we consider only the relevant registries (involving fcc- and hpc-hollow sites) and take the average of the computed core level shifts for the different C sites. This gives the red curve in Fig.~\ref{fig:cls_DFT}(a). The XPS spectrum is then built as a sum of Lorentzian functions (width 0.2 eV) centered in the shifts of each of the 200 C atoms in the cell. 
Note that since this method does not provide the absolute values of the energy shifts but rather their relative positions, we have aligned the peak of Gr/Ir(111) (dark green in Fig.~\ref{fig:cls_DFT}) with the experimental value.
The continuous distribution of C-Co distances results in a two peak XPS spectra (as shown in Fig.~\ref{fig:cls_DFT}c) that compares well with the experiment (Fig.~\ref{fig:cls_DFT}d).

In the same figure we have also plotted the calculated XPS spectrum of graphene on Ir. The core-level shift of graphene on Ir was computed using a commensurate slab of graphene on 4 layers of Ir, with a lateral periodicity of $4\sqrt{3}\times4\sqrt{3}$. The lattice parameter used (2.39~\AA{}) is an average value of the optimized lattice parameter for graphene (2.44~\AA{}) and Ir (2.34~\AA{}), in order to minimise the strain/stress on both systems (2\% both on Gr and on Ir). 
A full geometry relaxation was performed in a $\sqrt{3}\times\sqrt{3}$ cell resulting in an essentially flat graphene layer with a C-Ir distance of 3.45 \AA. We have also performed calculations for free standing graphene and for graphene on Co, presented in the SI, of the core level shifts as a function of the lattice parameter. We have used the value computed for free standing graphene, 2.25~eV/\AA{}, to correct the core-level shift for Gr@Ir, computed at a lattice parameter of 2.39~\AA{}, towards the value expected for a lattice of 2.44~\AA{}. 

So far we have considered an ideal moir\'e patterning (egg-box model), as described by Eq.~(\ref{eq:moire}), while the actual degree of corrugation is not only related to the lattice mismatched substrates, but also depends on the interaction strength of the Gr layer with the underlying metal surface. It is therefore reasonable to expect that the egg-box model underestimates the number of C atoms close to the surface. In fact, while this model describes well the corrugation of graphene on Ir(111) \cite{PhysRevLett.107.036101}, for the case of Co/Ir(111) the calculated C-Co interplane distances of about 2.0~\AA{} (as well as the charge density difference plots at fcc- and hcp-hollow sites reported in Ref. \cite{PhysRevB.87.041403}) highlight a stronger cobalt-carbon hybridization and the formation of a chemical bond. This interaction favours shorter Co-C distances, increasing the population of the valleys. This is coherent with STM results for graphene on Ni/Ir(111), in which a similar interaction is expected, showing that despite of a large corrugation of 1.51~\AA{}, 70\% of the C atoms are adsorbed with C-Ni distances between 2.0 and 2.2~\AA{}~\cite{Pacile2013}. This deviation from the ideal egg-box model explains both the smaller splitting of the two peaks in the computed spectra as well as the theoretical overestimation of the relative intensity for the lowest binding energy peak.

To summarize, according to our calculations the dependence of the chemical shifts on the C-Co distance has been found to be crucial in order to account for the XPS line-shape of the Gr/Co/Ir(111) interface. 
Even if the C-Co distance is determined by the registry of the C atoms with respect to the underlying Co surface, for the same  C-Co distance, the difference in the chemical shifts due to different registries is much smaller than the one due to the C-Co distances (see Fig. S2 in the SI).
The stretching of the underlying Co substrate as well as the Co-Ir hybridization have been tested to have a small effect of the C$1s$ core level shifts (See Fig. S1 in the SI).
Overall, the egg-box model used here reproduces well the corrugation of graphene on Ir with intercalated Co, giving the correct two peak splitting seen in the experimental XSP spectra.
This is interesting also in view of the possibility of depositing molecules, such as FePc, on the lattice resulting from the moir\'e pattern. In such a case, described in the following, the molecule-surface distance can be tuned via the corrugation of Gr, thereby influencing the electronic/magnetic coupling at the interface.

\subsection{FePc on Graphene with intercalated Co layer: adsorption sites and interaction}

The corrugated moir\'e superstructure of Gr, with an intralayer of Co, can drive the assembly of ordered superstructures of organo-metallic molecules\cite{BazarnikACSnano2013}. Metal phthalocyanines deposited on Gr/Ir(111) arrange in a nearly square lattice \cite{HamalainenJPCC2012}, driven by molecule-molecule interaction with a negligible molecule-surface mixing. Here Gr acts as a buffer layer, efficiently decoupling the molecules from the metallic substrate \cite{Scardamaglia2013,ScardamagliaLangmuir2013}. Single layer of Cobalt (Fe, Ni) intercalated between Gr and Ir, induces a moir\'e pattern with a pronounced corrugation and with a similar periodicity as Gr/Ir\cite{Pacile2014,PhysRevB.87.041403,Decker_JPhys_2014}. 

\begin{figure}
\begin{center}
\includegraphics[clip,width=0.4\textwidth]{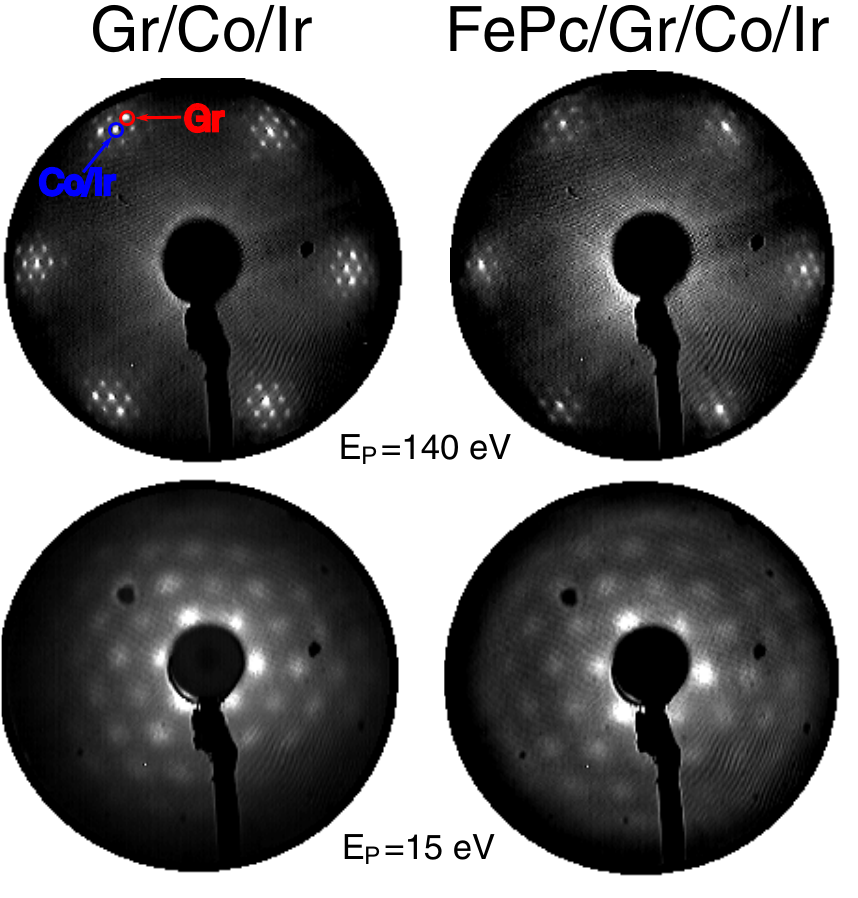}
\caption{\label{fig:LEED} LEED patterns of Gr/Co/Ir (left) and 0.4 ML FePc/Gr/Co/Ir (right) at 140 eV (upper panel) and 15 eV (lower panel) primary beam energy. The molecular adsorption does not alter the periodicity.
}
\end{center}
\end{figure}

The LEED pattern of Gr on Ir(111), with an intercalated ML of Co, presents well defined spots in an hexagonal structure, due to the Co intralayer, in registry with the underlying Ir(111) surface (Fig.~\ref{fig:LEED}, left panel). The moir\'e superstructure, due to the corrugated graphene, presents the same periodicity and symmetry after the Co intercalation, as clearly shown in the pattern recorded at low primary beam energy, sensitive to the graphene corrugated layer. The highly ordered moir\'e superstructure is preserved after the FePc adsorption (Fig.~\ref{fig:LEED}, right panel) and the LEED pattern shows spots with similar periodicity, with a slight diffuse background, due to lower ordering of the molecular domains.

Though the symmetry and periodicity of the moir\'e is unperturbed by the presence of FePc molecules, the surface potential for molecular adsorption is altered for the intercalated systems, compared with the bare Gr/Ir, and can strongly influence the molecular assembly and the molecule-substrate interaction ~\cite{BazarnikACSnano2013}. 

The adsorption geometry and the interaction strength of FePc molecules assembled on Gr/Ir(111), after the intercalation of about 1 monolayer (ML) of Co, can be investigated by studying first the role of the macrocycle and then the interaction of the central metal ion by means of absorption and core level photoemission spectroscopies. 

\begin{figure}
\includegraphics[width=0.5\textwidth]{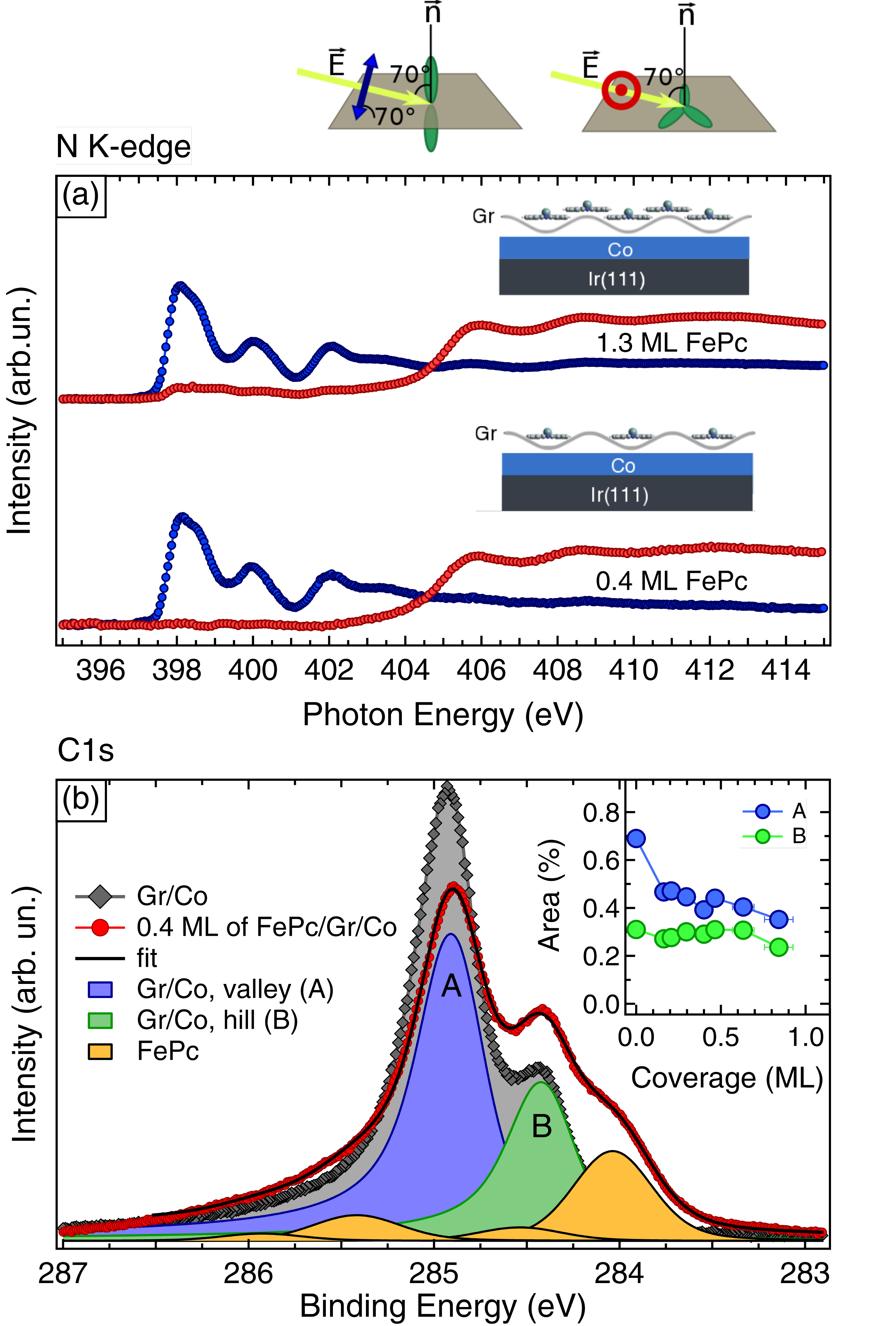}
\caption{\label{fig:C1s_NKedge} (a) NEXAFS spectrum of N K-edge for more (upper spectra) and less (lower spectra) than a FePc single layer adsorbed on Gr/1.5 $\pm$ 0.3 ML of Co. The red (blue) curve indicates the in-plane (out-of-plane) polarization, see sketches on top for clarity. (b) XPS spectrum of the C$1s$ core level before (grey diamonds) and after (red circles) the adsorption of 0.4 ML of FePc on Gr/1.0 $\pm$ 0.1 ML Co, together with the fitting curves and components (after deposition case). In the inset of panel (b) the evolution of the C$1s$ intensity of the Co-intercalated G/Ir(111) valley and top components, normalized with respect to the bare G/Co C$1s$ total area, at increasing molecular coverage is presented.}
\end{figure}

The absorption at the nitrogen K-edge, for 0.4 ML (1.3 ML) nominal thickness of FePc deposited on Gr/Co (Fig.~\ref{fig:C1s_NKedge}a), presents a first $\pi^{*}$-symmetry resonance at about 398 eV photon energy, composed by two unresolved structures due to photoexcitation from the N-isoindole and N-azomethine nitrogen atoms in the pyrrole ring to the LUMO. The higher $\pi^{*}$ energy structures at 400 and 402 eV rise from transitions to higher unoccupied states mainly localized on the pyrrole ring~\cite{Scardamaglia2013}. The highly dichroic N K absorption edge signal, with in-plane $\sigma^{*}$ states and out-of-plane $\pi^{*}$ levels, is independent on the FePc coverage, confirming adsorption of FePc molecules on Gr/Co in a flat lying configuration. Differently, direct adsorption of FePc on Co(001) films, epitaxially grown on Cu(001), unravels a significant molecular deformation due to the FePc interaction with the metallic substrate. The change in the $\pi^{*}$ region of the spectrum is due to a significant out-of-plane distortion of the molecular structure caused by the mixing with the underlying metallic states ~\cite{Klar_PRB_2013}. The presence of the graphene buffer layer in FePc/Gr/Co prevents the coupling of the pyrrole macrocycles with the underlying metallic layer, as observed also for FePc/Gr/Ir(111)~\cite{Scardamaglia2013}.

As clearly stated in the previous paragraph, the C$1s$ chemical shift is strongly influenced by graphene rippling, its spatial variation and the C-Co distances. A fine control of the C$1s$ lineshape, at increasing FePc molecular coverage, can unveil how the moir\'e  pattern can play an active role in the assembly formation. Indeed, in the C$1s$ XPS lineshape, the contribution from the Gr layer and the molecular components coexist (Fig.~\ref{fig:C1s_NKedge}b). The molecular contribution to the C$1s$ core-level has been accounted for with four Voigt components~\cite{MassimiJChemPhys2014, Papageorgiou_PRB_2003}, associated with the C atoms in the benzene (284.0 eV binding energy, BE) and pyrrole (285.4 eV) rings, together with the shake-up satellite due to transitions from the benzene HOMO to the LUMO (285.9 eV, the same excitations from the pyrrole ring are found to be of negligible intensity). The fourth component (284.5 eV) is attributed to a vibrational mode in the benzene rings. The energy difference between the molecular C$1s$ peaks and their relative intensities are in agreement with the parameters and the stoichiometry reported for a FePc thick film~\cite{MassimiJChemPhys2014}, suggesting the lack of significant charge transfer and/or molecular deformation for the pyrrole and benzene C atoms.

Looking at the contribution to the C$1s$ lineshape due to the Gr layer, the double structure (A and B), centered at 284.92 eV and 284.42 eV, corresponding to the Gr carbon atoms in valley(A)- and hill(B)-regions of the moir\'e superstructure\cite{Pacile2014}, is influenced by the presence of FePc adsorbed molecules. At low FePc molecular density, the C$1s$ component A shows a marked intensity reduction due to the FePc adsorption, while the component B stays almost constant. At higher FePc coverages, the C$1s$ B component is also reduced. The different intensity decrease of the two C$1s$ components, as a function of FePc molecular coverage (Fig.~\ref{fig:C1s_NKedge}(b) inset), suggests a hierarchical adsorption with preferential sites for the FePc molecules on the valley regions up to 0.7 ML. The trapping of FePc molecules in the valley regions can be favoured  by the modulation of surface dipole moments in the moir\'e cell due to the different registry between C and Co atoms. Indeed, the large molecular planar polarizability of FePc can be strongly influenced by the electrostatic potential gradient between the valley and the hill regions in the moir\'e superstructure~\cite{Dil2008}.  Site-selectivity has not been observed for FePc adsorbed on Gr/Ir(111) \cite{Scardamaglia2011}, where the slight corrugation across the moir\'e unit cell does not drive the molecular assembly and the FePc molecules arrange in a close-packed almost square lattice, where intra-molecular interactions dominate the assembly process.~\cite{Yang_JPCC_2012}. This interpretation is consistent with the theoretical analysis of the C$1s$ chemical shifts for Gr/Co as discussed in the previous section. Actually, also in the simpler Gr/Co case, our results indicate that the overall XPS line-shape is governed mostly by the dependence of the chemical shifts on the Gr-Co distance. Interestingly, here the moir\'e corrugation of Gr results in different distances between the FePc and the Co layer, originating a different electronic coupling.

\begin{figure}
\includegraphics[width=0.5\textwidth]{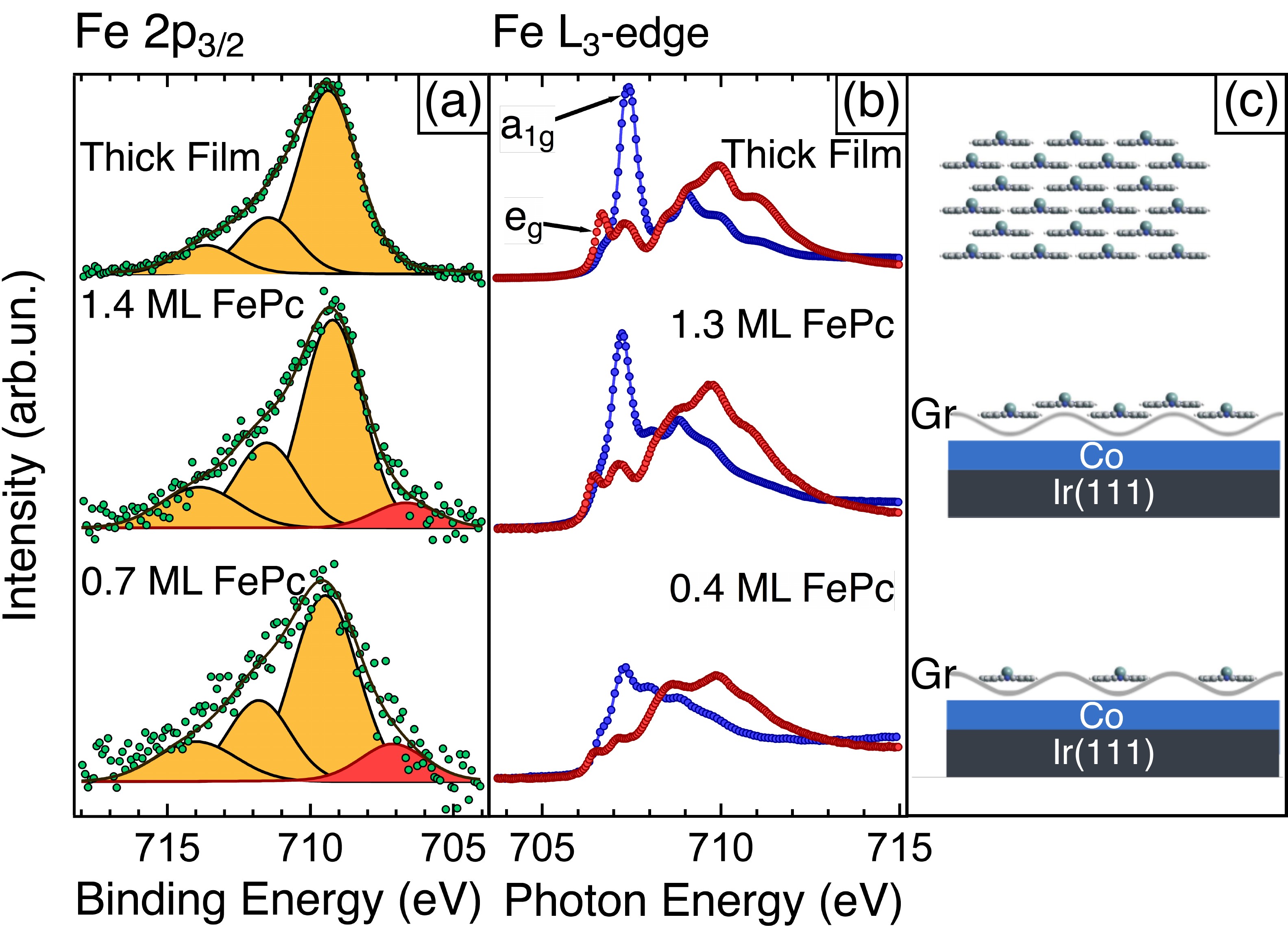}
\caption{\label{fig:Fe2p_Ledge} Fe$2p_{3/2}$ core level (a) and Fe L$_3$ absorption edge (b) spectra acquired for less and more than a single layer of FePc deposited on G/Ir(111) intercalated with 1.2 $\pm$ 0.2 ML and 1.5 $\pm$ 0.3 ML of Co, respectively. The XPS fitting curves are Gaussian peaks with 2.5 eV full-width at half-maximum (FWHM). The spectra for low FePc coverages, compared with the one of the TF (top spectrum) exhibit an extra component at low-binding energies, fingerprint of FePc-substrate interaction~\cite{MassimiJChemPhys2014}. The absorption associated with the $a_{1g}$ orbital (i.e. the one with a $d_{z^2}$ symmetry) for the FePc-TF (data from Ref.~\cite{GargianiPRB2010}), exhibits an intensity reduction for low molecular density. In panel (b) the red (blue) curve represents the in-plane (out-of-plane) polarization, as sketched at the top of Fig.~\ref{fig:C1s_NKedge}.}
\end{figure}

A selected set of core level photoemission spectra of Fe$2p_{3/2}$, as a function of FePc molecular density (sample sketches are presented in Fig.~\ref{fig:Fe2p_Ledge}c) up to the formation of a FePc thick film (TF), is shown in Fig.~\ref{fig:Fe2p_Ledge}(a). The core-level lineshape in the FePc-TF is characterized by a broad main peak centered at 709.5 eV BE, whose asymmetry can be associated to the presence of multiplet structure in the photoemission final states of the central Fe$^{+2}$ ion, as observed also for Fe-phthalocyanine and Fe-porphyrin molecules~\cite{SchmidJPCC2011, MassimiJChemPhys2014, Isvoranu_JPCC_2011}. Several approaches have been used for evaluating the multiplet structure of the Fe$2p$ core-level, either applying a Zeeman-like final state effect, based on the four split levels $m_j$ = 3/2, 1/2, -1/2, -3/2 components of the $2p_{3/2}$ state~\cite{Isvoranu_JPCC_2011}, or in split states based on their total angular momentum ($J$ = 5/2, 3/2, 1/2) \cite{SchmidJPCC2011}. The adopted fit procedure of the Fe$2p_{3/2}$ core-level for the thick FePc film, using a Zeeman-like analysis, is a simplified model where the four $m_j$ sublevels are partially unresolved and modeled using two peaks centered at 709.5 and 711.5 eV and a broader satellite peak at 714.0 eV. A further Fe$2p_{3/2}$ component at low FePc molecular density has been introduced. It is noteworthy that, independently of the chosen fitting model, an extra peak is present when comparing the FePc supra-monolayer and sub-monolayer with the thin film. For 0.7 ML FePc, i.e. when all the valley regions are filled, the extra component, located at 706.5 eV, has an area equal to the 10\% of the total intensity and the remaining signal is associated to the Fe(II) multiplet features. The peak at lower binding energy in the XPS Fe$2p_{3/2}$ data, for sub-monolayer FePc molecular density, can be associated to FePc molecules directly in contact with the underlying Gr valley regions, suggesting a charge transfer and orbital mixing from the substrate to the central Fe ion\cite{SchmidJPCC2011, MassimiJChemPhys2014, Isvoranu_JPhys_2010}.
It is worth noting that FePc deposited on the gently-corrugated Gr/Ir(111) superstructure does not present any extra-component due to the Fe ion interaction with the underlying substrate (as discussed in the SI), confirming the decoupling role of Gr, as a buffer layer, when the MPc are deposited on a low corrugated moir\'e with a dominant molecule-molecule interaction~\cite{Scardamaglia2013, Scardamaglia2011}. A different description was recently proposed for FePc deposited on Gr grown on Ni(111) by Uihlein et al.\cite {Uihlein_2014}. Here, the authors infer the tail of the Fe$2p_{3/2}$ component at about 707 eV to a direct contact of Fe and Co atoms in small uncovered regions at surface. In the present study, on the basis of several crossed measurements, we can rule out the presence of holes in the Gr layer grown on Ir(111), as well as a partial intercalation of Co atoms, which both would put the molecules in direct contact with the intercalated layer.

Instead, the presence of an extra-component due to a charge transfer from the metallic substrate to the central Fe atoms has been observed for other FePc-metal interfaces~\cite{MassimiJChemPhys2014, SchmidJPCC2011, BaiJPCC2008}. The Fe$2p$ signal, measured for FePc molecules adsorbed on Au(111), presents a similar asymmetric lineshape, due to an interaction of the FePc molecules with the Au(111) surface~\cite{SchmidJPCC2011}. A higher interaction strength, i.e. a dominant contribution arising from the low-binding energy structure, of Fe ion and the underlying surface is observed for the Fe$2p$ signal measured for FePc molecules adsorbed on Au(110)~\cite{MassimiJChemPhys2014} and Ag(111)~\cite{BaiJPCC2008}, suggesting a significant charge transfer from the metal surface. The lower intensity of the extra-component, observed for FePc adsorbed on Gr/Co/Ir, suggests a lower charge transfer to the Fe ion metal center compared to the bare metal case.

Further FePc deposition induces a decrease in the intensity of the extra component (at 1.4 ML FePc it accounts for the 5\% of the total intensity). These results confirm that the FePc-Co intralayer coupling varies between the hill and valley regions of the moir\'e pattern. Indeed the reduction of the extra-component intensity, at higher FePc coverage, unveils a lower interaction between Fe and Co when the molecules are not in direct contact with the Gr-Co interface. 

A further confirmation of a direct interaction between the Co intralayer and the Fe metal ion centers can be obtained by measuring the absorption at the Fe L$_{2,3}$-edges of the FePc/Gr/Co system. A selected set of Fe-L$_3$ edges, as a function of FePc coverage, are shown in Fig.~\ref{fig:Fe2p_Ledge}(b). The L$_{2,3}$ edges show a dichroic response, confirming the molecular flat-lying configuration. The electrons excited from the $2p_{3/2}$ or $2p_{1/2}$ initial states induce a multiplet structure. Considering the symmetry, the different main absorption features can be assigned taking into account the ligand field splitting of the Fe-related molecular states, the sequence of spin-split molecular orbitals, and the final spin state ($S$=1). The main absorption features, for a FePc molecular coverage higher than a ML, reflect the attribution of the main features to the sequence of molecular orbitals for an isolated FePc molecule or a FePc film~\cite{BettiJPCC2010}. In the out-of-plane polarized spectrum, (Fig.~\ref{fig:Fe2p_Ledge} (b) top blue curve), the excited electrons accede to states localized normal to the molecular plane, and therefore the main L$_3$ peak at 707.5 eV photon energy is due to the a$_{1g}$ ($d_{z^2}$) empty state and the structure at 706.5 eV can be attributed to the e$_g$ ($d_{xz,yz}$) states. The Fe L$_{3}$-edge for the FePc sub-ML on Gr/Co maintain the same dichroic in-plane/out-of-plane response (owing to their flat-lying orientation), but with a slightly different lineshape in the near-edge region, with respect to the FePc at higher molecular density. The out-of-plane polarized spectra, sensitive to the molecular levels with a dominant symmetry perpendicular to the FePc plane, show stronger variations. In particular, at the lowest FePc density (Fig.~\ref{fig:Fe2p_Ledge}(b), bottom blue curve) the a$_{1g}$ peak intensity is strongly reduced, as also the pre-edge peak at 706.5 eV, attributed to the e$_g$ state. In contrast, the multiplet features at higher photon energy, related to molecular states with a dominant in-plane b$_{1g}$ symmetry, show slight variations with respect to the molecular film. The evolution of the molecular levels as a function of FePc density confirms an intermixing of the out-of-plane molecular states with the underlying Co intralayer metal states, dominant when the FePc are in the valley region, where the Gr-Co distance is reduced.
Similarly, FePc adsorption on Au(110) suggests a charge transfer from gold to the out of plane molecular states and the electronic structure of the filled and empty states is strongly altered~\cite{GargianiPRB2010,BettiJPCC2010}. On the other hand, when the moir\'e corrugation is less pronounced and the Gr is less interacting with the metallic substrate, as observed for Gr grown on Ir(111), the FePc are almost unperturbed by the presence of the Ir(111) surface\cite{Scardamaglia2013,ScardamagliaLangmuir2013}.

According to our findings, in the FePc/Gr/Co system, the molecular assembly and the molecules-substrate interaction is strongly influenced by the moir\'e rippling. Indeed, the XPS and NEXAFS experiments unveil a partial intermixing and hybridization when the FePc molecules are trapped in the valley regions of the graphene moir\'e superstructure, closer to the underlying Co, while a negligible interaction is revealed  at higher FePc coverage, when the molecules start to adsorb in the hill region. The trapping of the FePc molecules in the valley region can be induced by ($i$) the modulation of the surface dipole of the rippled graphene due to the different registry and ($ii$) by different vertical distance of the graphene C atoms and the Co intralayer sites. The gradient of these two parameters, between the valley and the hill regions, drives the (de)coupling of the FePc metal centers with the underlying Co metal.  

\section{Conclusions}
In this work we have investigated the interaction mechanism of the FePc molecule on the artificially corrugated Gr/Co interface. We first focused on the Gr/Co template in two different registries, flat (1$\times$1) Gr/Co and rippled Gr/Co/Ir(111), by comparing C$1s$ photoemission results with {\it ab initio} calculations based on DFT. Concerning the flat Gr/Co interface, we show how the two non-equivalent adsorption sites of C atoms in the top-fcc registry give a small splitting of the corresponding components in the C$1s$ peak. Instead, the vertical distance of the C atoms from the Co surface, ranging in our model from 2.05 to 3.30 ~\AA{}, yields a much more pronounced shift. On the basis of these results, we have constructed an ideal Gr/Co moir\'e superstructure using the egg-box model. We show that the rigid model describes the main features of the C$1s$ peak of the Gr/Co/Ir interface, demonstrating that the chemical shift of C$1s$ is mainly governed by the vertical distance of C atoms. 

Accordingly, the interaction of FePc molecules comes out to be different within the rippled superstructure. By XPS and NEXAFS measurements, we provide evidence that, upon FePc molecular deposition on the Gr/Co/Ir interface, the valley regions are first filled, and an hybridization of Fe-related out of plane orbitals with the underlying metal surface has been detected. At higher FePc molecular densities, the hills of the moir\'e superstructure begin to be covered, the mixing of the Fe orbital is negligible and the molecule-molecule interaction becomes dominant. Further, our experiment suggests that a fine control of the rippling and of the surface potential of the Gr template is required in order to drive not only the self-assembly of supramolecular network, but also to tune the hierarchical interaction between molecular units and towards the Gr-metal template. 

\section{Methods}\label{sec2:methods}

\subsection{Experimental details}

Near-Edge X-ray Absorption Fine Structure (NEXAFS) measurements were performed at the BOREAS beamline of the Alba synchrotron radiation facility (Barcelona, Spain), in a Ultra-High-Vacuum (UHV) chamber with a base pressure in the low $10^{-10}$ mbar regime. The data were collected by measuring the sample drain current, excited with linearly (horizontally and vertically) polarized radiation impinging on the sample at grazing incidence ($\theta_i$=70$^\circ$). The NEXAFS spectra were collected in the two opposite orientations of the scattering plane with respect to the electric field of the linearly polarized X-ray beam, namely, transverse magnetic (TM) and transverse electric (TE) polarization. The polarization condition was changed by modifying the phase of the Elliptically Polarizing Undulator (EPU); thus, TE polarization corresponds to the electric field oriented in the surface plane, while TM corresponds to an orientation of the electric field with an angle of 70$^{\circ}$ with respect to the sample surface (see inset in Fig.~\ref{fig:C1s_NKedge}).

The core level (Fe$2p$) spectra were acquired in the LoTUS surface physics laboratory (Sapienza University of Rome), with Mg K-$\alpha$ radiation (1256.6 eV) and collecting the photoemitted electrons with a VG Microtech Cam-2 hemispherical analyzer with an overall resolution of 1 eV.
The C$1s$ XPS measurements were performed at the SuperESCA beamline (checked at APE beamline) of the Elettra synchrotron light source (Trieste, Italy) where photoelectrons were collected with a Phoibos electron energy analyzer equipped with a homemade delay line detection system. All the XPS spectra were acquired at room temperature and in a normal emission geometry.

All laboratories are equipped with analogous ancillary facilities for sample preparation and molecular deposition and control. Surface quality and cleanness were checked by means of Low Energy Electron Diffraction (LEED) and/or core level photoemission. The Ir(111) surface was prepared with repeated cycles of sputtering (Ar$^+$ with kinetic energy of 2000-2500 eV) followed by annealing (above 1300 K). The Gr sheet was grown on the Ir(111) surface following a Temperature Programmed Desorption (TPD) procedure: the surface was exposed for 120 s to $5\times 10^{-7}/2\times 10^{-8}$ mbar of ethylene (C$_2$H$_4$) and subsequently annealed over 1300 K for 60 s, this procedure was repeated up to the completion of the single layer Gr, checked by monitoring the C$1s$-Ir$4f$ core levels intensity ratio.

Metallic Co was evaporated by an electron beam source and deposited on the Gr/Ir(111) system. The Co coverage has been determined evaluating the Ir$4f$ core level attenuation and checking the lineshape of Gr C$1s$ after intercalation ~\cite{Pacile2014,VitaPRB2014}, at the LoTUS and Elettra laboratories. At the BOREAS beamline, the Co coverage was calibrated with Auger Electron Spectroscopy and then confirmed by evaluating the jump-edge ratio~\cite{ArvanitisSpringer1996}. Finally, Co intercalation was obtained by annealing the sample to 500-800 K for several minutes, according to Ref.~\cite{Pacile2014, VitaPRB2014, PhysRevB.87.041403}. After the intercalation of Co, the LEED presents the moir\'{e} superstructure pattern, up to about four Co layers.

The commercially-available AlphaAesar FePc molecules were deposited by using a home-made resistively heated quartz crucible. The FePc evaporation rate was 0.30$\pm$0.05\AA/min, measured with a quartz-crystal-microbalance considering a molecular film density of 1.52 g/cm$^3$~\cite{Singh2010}. The estimated nominal thickness, with an error of the 15\%,  was double-checked, at LoTUS and Elettra laboratories, by evaluating the Ir$4f$ further attenuation and by comparing the N $1s$ core-level (not shown) intensity at each growth step.

\subsection{Computational details}

Density functional theory calculations were performed using the 
Quantum ESPRESSO~\cite{gian+09jpcm} package.
The local density approximation (LDA, Perdew-Zunger parametrization\cite{perd-zung81prb}) was adopted together with plane wave basis set and norm-conserving pseudo-potentials to model the electron-ion interaction. The kinetic energy cutoff for the wave functions was set to 75 Ry. Specifically, the XPS calculations for graphene on Co are performed considering a commensurate 4$\times$4 supercell of graphene on top of a Co(0001) slab with four layers of Co atoms. 
The Brillouin zone was sampled by using a 4$\times$4$\times$1 $\mathbf{k}$-points grid.
The supercell was checked (up to 8$\times$8 for free-standing graphene) to be large enough to accomodate a single C pseudopotential with a full $1s$ core-hole avoiding spurious interactions with its replica. 
The (relative) chemical shifts of C$1s$ levels for non-equivalent C atoms were computed~\cite{trig+98prb} as total energy differences between systems with the same geometries with and without core-hole inclusion.
The core level shifts were computed for different distances between graphene and the Co surface. Different registries were also considered.

\section{Acknowledgements}
This work has been supported by the European Science Foundation (ESF) under the EUROCORES Program EuroGRAPHENE. Partial support has been provided by the Centre of Excellence MaX -- MAterials at the eXascale -- (grant no. 676598) and by Sapienza University of Rome. The authors thank  the SuperESCA, APE (Elettra-Italy) and BOREAS (Alba-Spain) beamlines staffs for their experimental support and Marco Papagno for the experimental collaboration during the APE beamtime. Computational resources were provided by the PRACE project on the FERMI machine at CINECA (Grant No. Pra11\_2921).


\newpage
\bibliographystyle{achemso}
\bibliography{biblio_NEXAFS}

\end{document}